\documentclass[a4paper,
keeplastbox,   
]{jacow}
\usepackage{pdfpages,multirow,ragged2e} %
\usepackage{xcolor}

\makeatletter%
\ifboolexpr{bool{xetex}}
{\renewcommand{\Gin@extensions}{.pdf,%
		.png,.jpg,.bmp,.pict,.tif,.psd,.mac,.sga,.tga,.gif,%
		.eps,.ps,%
}}{}
\makeatother

\ifboolexpr{bool{xetex} or bool{luatex}} 
{}                                      
{\usepackage[utf8]{inputenc}}           

\usepackage[USenglish]{babel}

\ifboolexpr{bool{jacowbiblatex}}%
{%
	\addbibresource{jacow-test.bib}
	\addbibresource{biblatex-examples.bib}
}{}
\begin{document}
	
	\title{Comparative Evaluation of Xilinx RFSoC Platform for Low-Level RF Systems}

	\author{S.D. Murthy\thanks{sdmurthy@lbl.gov}, V. Moore, Q. Du, A. Jurado, \\
			K. Penney, M. Chin, D. Nett, B. Flugstad, LBNL, Berkeley, CA 94720, USA}

	\maketitle

	\begin{abstract}
		The rapid advancement of Radio Frequency System-on-Chip (RFSoC) technology from Xilinx (AMD) has enabled the integration of high-speed data converters and programmable logic within a single package. RFSoC platforms are already widely adopted in telecommunications, radar, and satellite communications, where they promise reductions in system footprint and power consumption. However, their suitability for Low-Level RF (LLRF) control systems in accelerator environments — where stability requirements are critical — has not been quantitatively evaluated. This paper presents a comparative measurement-based assessment of RFSoC-based and conventional LLRF designs, focusing on signal fidelity, phase noise, latency, system complexity, and integration challenges. The advantages and challenges of adopting RFSoC-based direct conversion architectures are discussed, providing guidance for future LLRF system implementations.
	\end{abstract}
	
	\section{INTRODUCTION}
	As next generation particle accelerators evolve toward high beam currents, smaller emittance, and tighter tolerances, the requirements on the LLRF control systems have become increasingly stringent~\cite{Nature}. Modern accelerators dictate sub-0.01$^\circ$ in phase stability with amplitude regulation of $0.01\%$, pushing the limits of conventional LLRF architectures in terms of latency, system complexity, and noise performance.
	
	In parallel, the semiconductor industry has introduced highly integrated platforms such as Xilinx's RFSoC, which combines portions of traditional RF-frontend including mixers, high-speed data converters (ADCs and DACs), and programmable logic in a single package. These devices are already transforming fields such as telecommunications, radar, radio astronomy, and satellite communications by enabling up to sixteen transmit and receive channels, facilitating advanced Multiple-Input Multiple-Output (MIMO) architectures, and thereby significantly increasing the channel density while reducing power consumption and footprint.
	
	Existing deployed LLRF systems typically employ RF signal-conditioning chain, consisting of attenuators, amplifiers, and monitoring circuits, followed by a RF mixing stage to down-convert the RF signal to an Intermediate frequency (IF) suitable for moderate speed ADCs. The digitized IF signal is then processed in a FPGA to implement feedback control and other diagnostic functions. Finally, the processed signal is converted back to RF through an up-conversion stage and injected back into the accelerator system~\cite{LCLS-II}. This conventional approach, while proven, requires many discrete analog components, additional frequency conversions, careful calibration, and a large number RF coaxial cables, all of which adds to the system complexity and increase in the potential failure points. RFSoC technology presents as an attractive option for LLRF systems, particularly for direct-conversion architecture~\cite{RFSoC} by bypassing the complex analog front-end.
	
	\section{RFS\MakeLowercase{o}C Architecture}
\begin{figure}[!htb]
		\centering
		\includegraphics*[width=1\columnwidth]{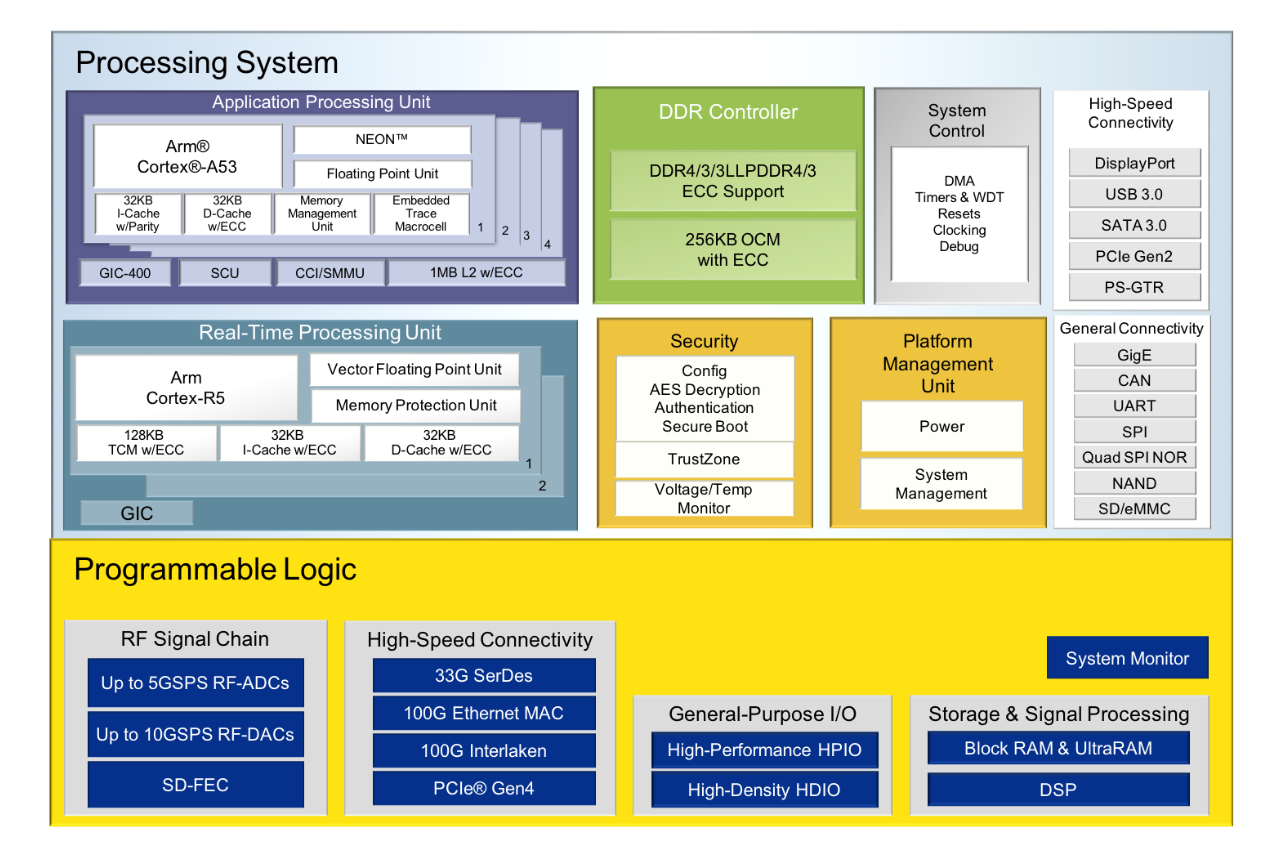}
		\caption{Zynq UltraScale+ RFSoC top-level architecture.}
		\label{fig:rfsoc_arch}
	\end{figure}

	The RFSoC architecture is centered around a highly integrated data path between on-chip RF data converters and the Programmable Logic (PL), enabling direct RF sampling without IF conversion stages~\cite{RFSoC2}. The embedded ARM Cortex‑A53 application processors and Cortex‑R5 real‑time processors provide system‑level control, configuration, and coordination with the PL, as shown in Figure~\ref{fig:rfsoc_arch}.
	
\begin{figure}[!htb]
	 	\centering
	 	\includegraphics*[width=1\columnwidth]{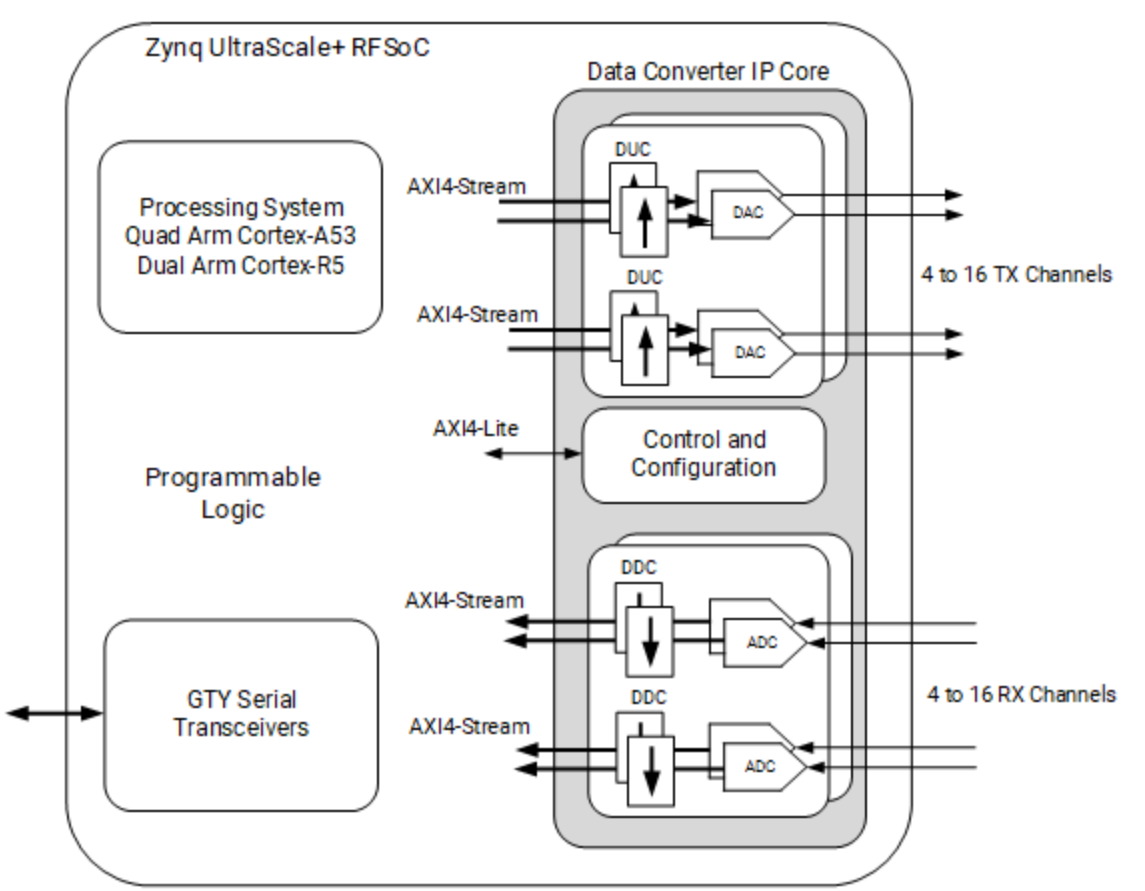}
	 	\caption{Zynq UltraScale+ RFSoC PL data path.}
	 	\label{fig:rfdc}
	 \end{figure}

	 Each data converter tile contains multiple channels (typically one, two, or four depending on the generation), on-chip Digital Up-Conversion (DUC) or Digital Down‑conversion (DDC) blocks, programmable complex digital mixers with Numerically Controlled Oscillators (NCOs), and decimation or interpolation filters, with outputs connected to the fabric through a dedicated low‑latency Advanced eXtensible Interface (AXI) Stream interface. The architecture supports both flexible clocking network with internal on‑chip PLLs for each tile and external references, with provisions for Multi‑Tile Synchronization (MTS) to align the relative latency of all ADC and DAC channels across multiple tiles when enabled. MTS ensures phase‑coherent sampling and output by synchronously resetting internal dividers and aligning FIFO pointers using a common reference. While MTS removes channel-to-channel timing differences, deterministic latency fixes the absolute end-to-end delay through the device across resets and power cycles by adjusting all the tiles to a defined target latency. Together, these built-in synchronization capabilities enhance LLRF system performance by providing sample-to-sample synchronization critical for maintaining sub-degree phase stability and precise amplitude regulation, provided that PCB clock and signal traces are delay‑matched. In addition, it integrates serial transceivers that provide high‑bandwidth connectivity to external systems for real-time data streaming and precise timing interfaces as shown in Figure~\ref{fig:rfdc}.
	 
	 Focusing on the ZCU208 evaluation platform~\cite{ZCU208} for the remainder of this paper, the board features a Generation\thinspace 3 device with ZU48DR FPGA integrating eight 14‑bit ADC channels capable of sampling at up to 5\thinspace GHz and eight 14‑bit DAC channels with sampling at up to 10\thinspace GHz. This multi‑channel configuration makes the ZCU208 an ideal platform for prototyping and evaluating direct‑sampling LLRF architectures in a controlled laboratory environment. All configuration, control, and data acquisition in this work are implemented using the open-source PYNQ framework~\cite{PYNQ}, enabling rapid development through Python APIs and overlays.

	 \section{EVALUATION AND COMPARISON}
	 The ZCU208 RFSoC device was evaluated to extract key performance metrics including phase noise, DAC and ADC characterization, crosstalk, and latency with all measurements performed at a sampling frequency of 4\thinspace GHz. Careful design considerations in the front‑end conditioning signal chain are essential to minimize any potential contributions.

	 \subsection{Phase Noise}
	 Phase noise refers to random fluctuations in the phase of a signal, typically expressed in dBc/Hz at a specified frequency offset from the carrier. In LLRF systems, the phase noise of the ADC and DAC sampling clocks directly impacts the system’s ability to regulate the amplitude and phase of the cavity field, as clock instability introduces amplitude and phase errors into the digitized cavity probe signal and the generated drive signal. Low phase noise sampling clock is therefore critical to achieving the sub‑degree phase stability and precise amplitude control required by modern accelerators.

	 The RFSoC family of evaluation boards are equipped with CLK104 RF clock add‑on board, which provides differential reference clocks for the data converters along with the synchronization clocks required to achieve MTS. These are generated using two on‑board PLL devices: the LMK04828 (a jitter cleaner and frequency divider) and the LMX2594 (a wideband synthesizer)~\cite{CLK104}. The ZCU208 platform was evaluated under three distinct clocking configurations to quantify the phase noise contribution from the sampling clock path:
	 \begin{itemize}
		\item Stock CLK104 board: This configuration employs the LMK04828 in jitter‑cleaner mode, which accepts a 500\thinspace MHz reference clock input and processes it into a stable, cleaned output. This cleaned clock signal is then provided to the LMX2594 synthesizer, which generates high-frequency sampling clocks for driving the ADC and DAC. This represents the default clocking scheme supplied with RFSoC evaluation kits.

		\item Modified CLK104 board: Reworked CLK104 board to configure LMK04828 to operate purely in distribution mode, significantly reducing its jitter contribution. The LMX2594 loop filter components were also updated based on simulation results to improve synthesizer phase noise.

		\item Internal on‑chip PLLs: Sampling clocks are generated directly by the RFSoC’s on‑chip tile PLLs from an external differential reference, allowing for the potential to bypass the CLK104 board. While this setup can simplify the signal path, it may introduce higher inherent jitter compared to optimized external clocking solutions.
	 \end{itemize}
	 
\begin{figure}[!htb]
	 	\includegraphics*[width=1\columnwidth]{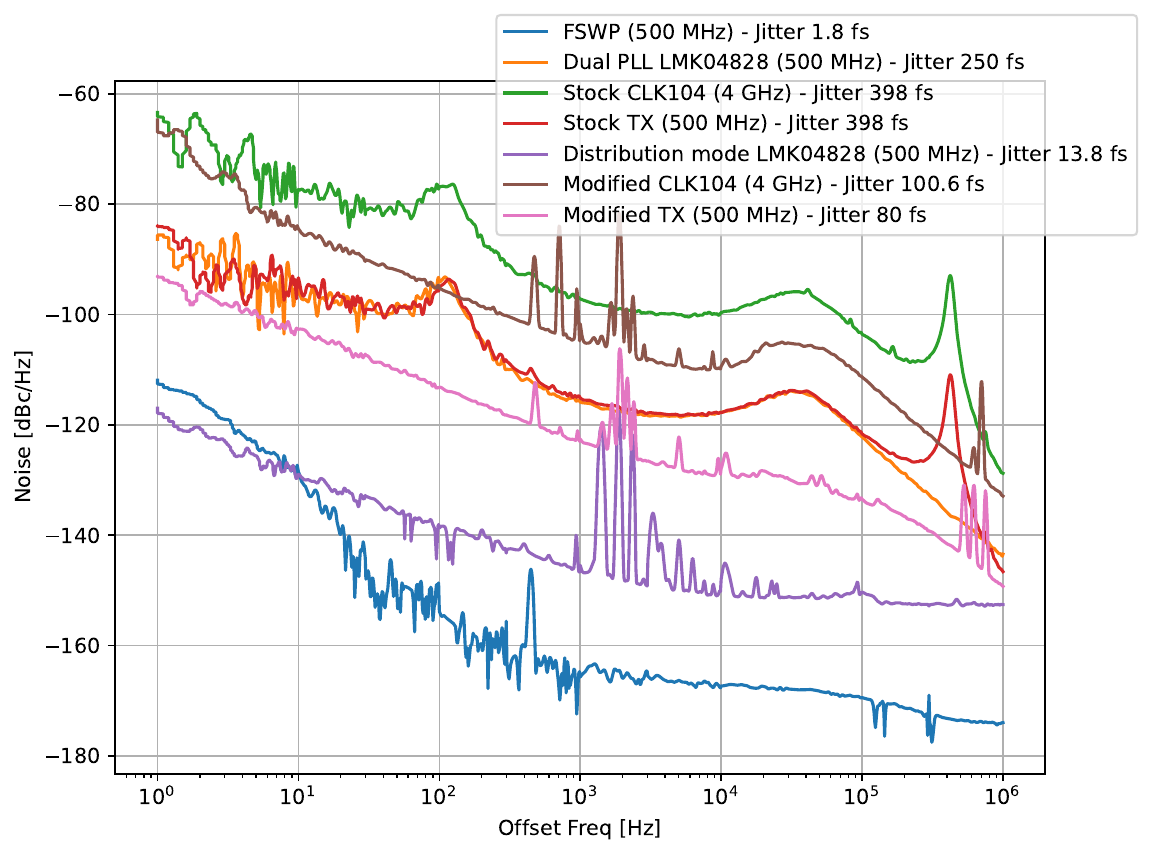}
        \caption{Measured phase noise with different configurations.}
        \label{fig:phs_noise}
	 \end{figure}

	 Figure~\ref{fig:phs_noise} shows the integrated phase noise of different clocking configurations in 1\thinspace Hz to 1\thinspace MHz offset range, measured using Rohde and Schwarz FSWP phase noise analyzer. The blue trace represents the additive phase noise of the FSWP, which serves as the 500\thinspace MHz reference for the system, yielding an integrated time jitter of 1.8\thinspace fs, establishing a baseline noise floor. In the stock CLK104 configuration, the LMK04828 in jitter‑cleaner mode contributes approximately 250\thinspace fs of jitter, making it about 18× higher than the modified configuration (purple trace, 13.8\thinspace fs) and dominates the total clock noise. The green trace is the absolute phase noise of the stock CLK104 at 4\thinspace GHz sampling frequency, achieving 398\thinspace fs integrated jitter; however, with the LMK04828 in distribution mode and optimized LMX2594 loop filters (brown trace), the time jitter is 100.6\thinspace fs. In the final transmit path (CLK104 + DAC), the measured additive phase noise with stock CLK104 is 398\thinspace fs but with modified CLK104 it is approximately 80\thinspace fs. This comparison highlights substantial improvements in jitter performance achieved by optimizing the clocking architecture through careful refinement of the PLL configuration and loop filter parameters, which in turn reduced the LMK04828's jitter contribution and improved sampling clock phase noise. Future work will focus on the direct injection of sampling clocks into the ADCs and DACs, which may bring the system closer to the 30\thinspace fs phase noise performance benchmark of traditional LLRF systems~\cite{ALSU} by minimizing jitter and simplifying the clock path.

	 \subsection{DAC and ADC Characterization}
	 The performance of the integrated DAC and ADC in the ZCU208 RFSoC was characterized to assess their suitability for accelerator LLRF control. For wideband dynamic tests, a single‑tone inputs/outputs were used to determine Signal‑to‑Noise ratio (SNR), Spurious‑Free Dynamic Range (SFDR) and Noise Spectral Density (NSD).

\begin{figure}[!htb]
	 \includegraphics*[width=1\columnwidth]{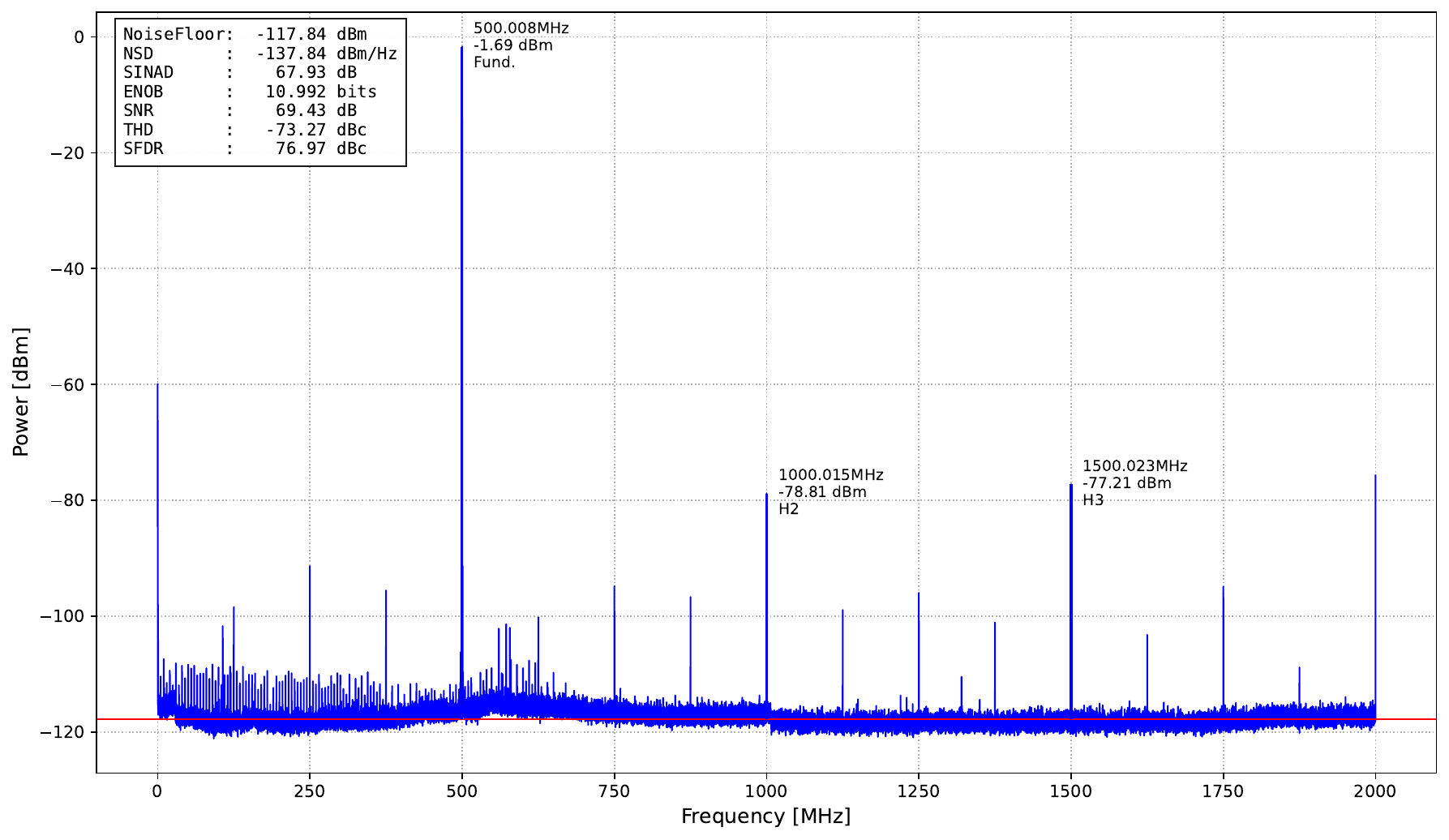}
	 \caption{DAC spectrum output at 500\thinspace MHz.}
	 \label{fig:dac_spectrum}
	 \end{figure}

	 For the DAC characterization, multiple single tone frequencies were generated using the built-in mixer in both first and second Nyquist zones, with an interpolation factor of 8. Figure~\ref{fig:dac_spectrum} shows the DAC spectrum output at 500\thinspace MHz tone frequency measured using the FSWP spectrum analyzer with a complete Nyquist bandwidth. These wideband measurements yielded a SNR of 69.43\thinspace dB, SFDR of 76.97\thinspace dBc, NSD of -137.84\thinspace dBm/Hz, and ENOB of 10.992\thinspace bits. Considering a closed-loop LLRF controller bandwidth of 200\thinspace kHz, the DAC's SFDR exceeded 88\thinspace dBc, which is close to the conventional LLRF system~\cite{ALSU}.

\begin{figure}[!htb]
	 \includegraphics*[width=1\columnwidth]{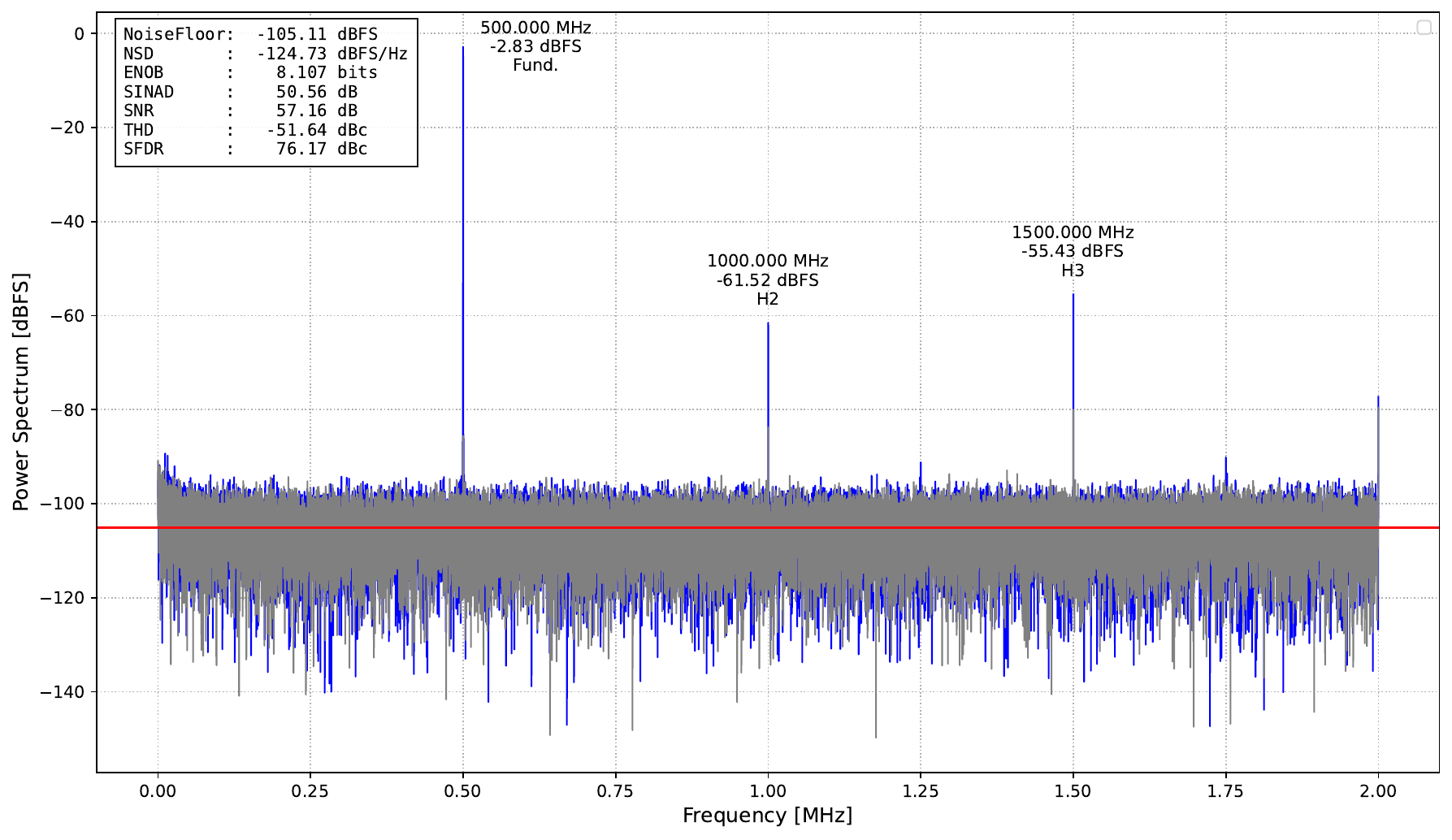}
	 \caption{ADC power spectrum at 500\thinspace MHz.}
     \label{fig:adc_spectrum}
	 \end{figure}

\begin{figure}[!htb]
	 \includegraphics*[width=1\columnwidth]{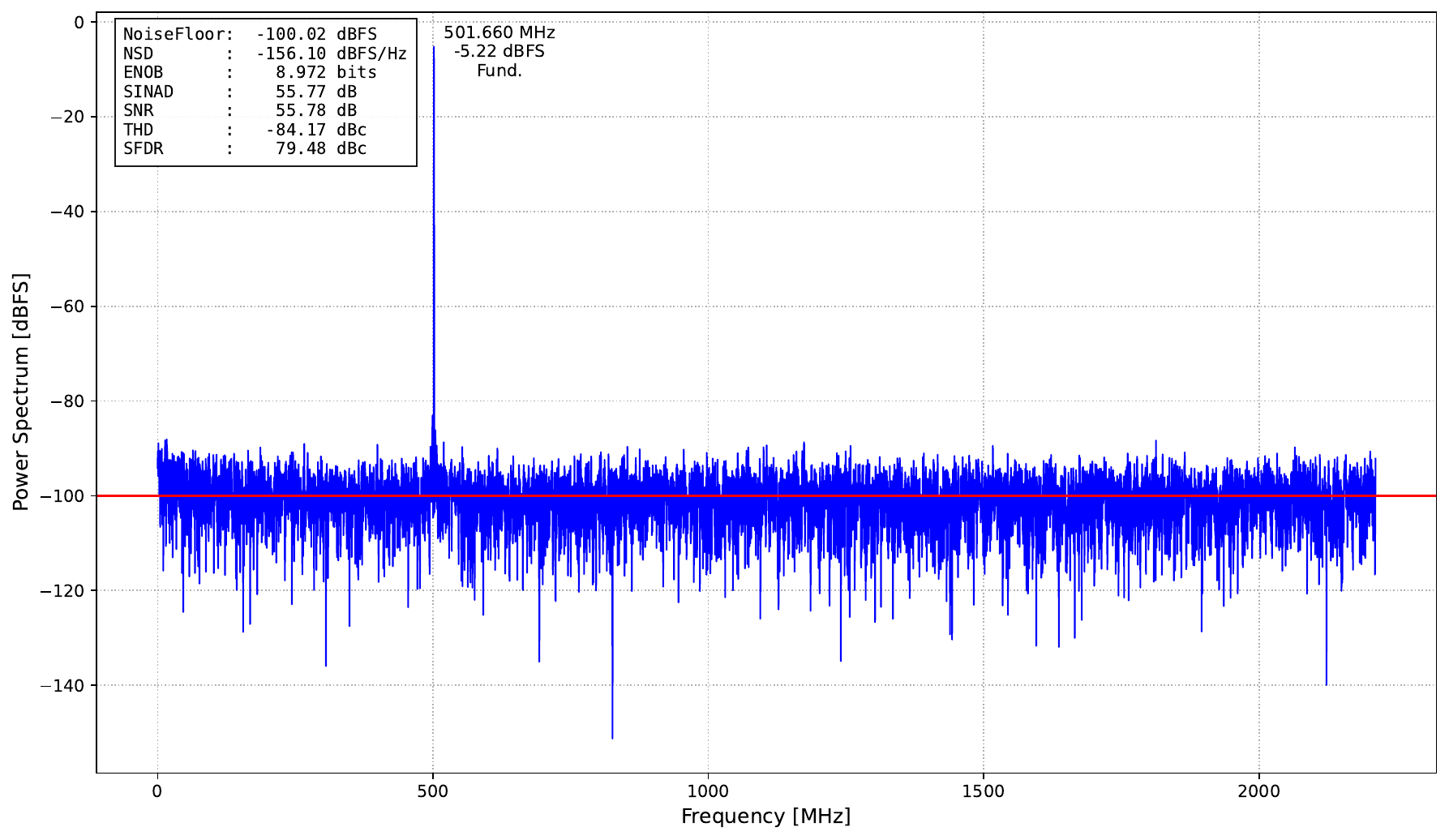}
	 \caption{Xilinx evaluation tool ADC spectrum at 501.66\thinspace MHz.}
	 \label{fig:adc_spectrum_ev}
	 \end{figure}

	 To characterize the ADC, an external RF synthesizer with a narrow band-pass filter was used to generate a single tone frequency at 500\thinspace MHz. Figure~\ref{fig:adc_spectrum} compares th ADC power spectrum without any signal and with the tone input, with no built‑in decimation factor enabled. The spectrum without a input signal shows harmonics at multiples of sampling frequency (500\thinspace MHz, 1\thinspace GHz, and 1.5\thinspace GHz), indicating a need for further investigation. To further access the ADC performance, a set of ADC data was captured using Xilinx's evaluation tool by looping back the DAC output with the same band-pass filter. The resulting power spectrum is shown in Figure~\ref{fig:adc_spectrum_ev}, with no significant harmonics or spurs observed, and reporting a SNR of 55.78\thinspace dB, SFDR of 79.48\thinspace dBc, NSD of -156.10\thinspace dBFS/Hz, and ENOB of 8.972\thinspace bits.
	 
	 \subsection{Crosstalk}
\begin{figure}[!htb]
	 \includegraphics*[width=0.9\columnwidth]{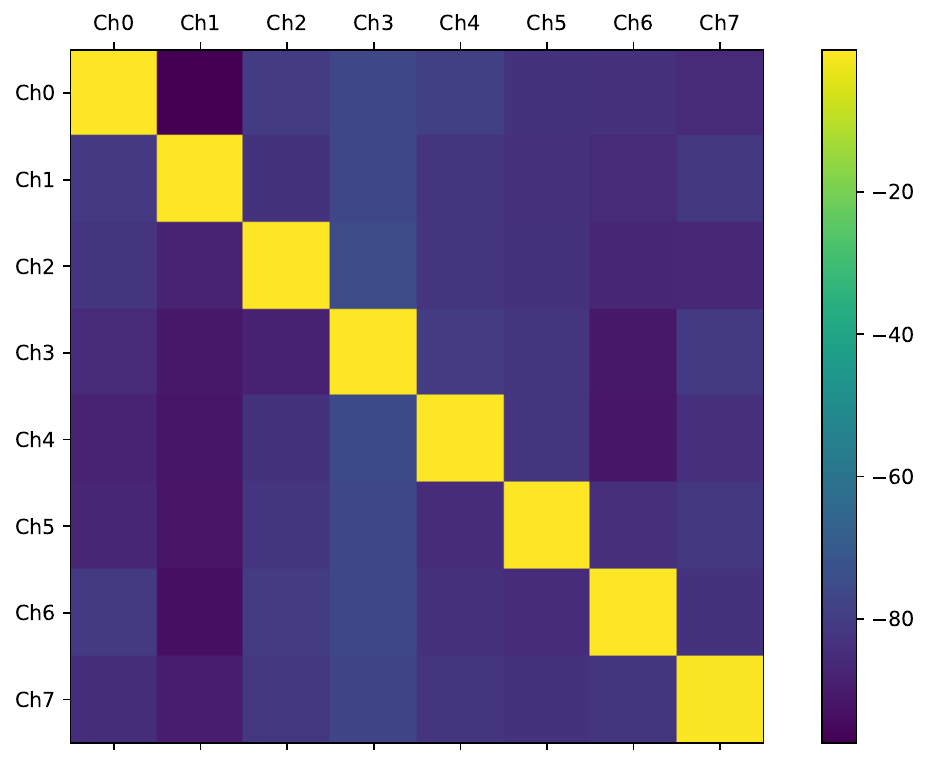}
	 \caption{Receiver channel-to-channel isolation at 500\thinspace MHz.}
	 \label{fig:crosstalk}
	 \end{figure}

	 In multi‑channel RF systems, crosstalk refers to the undesired coupling of signals between channels, which degrades signal fidelity and can distort measurements. For LLRF applications, low crosstalk is critical because leakage between adjacent channels can introduce phase and amplitude errors, reducing overall system stability. For the ZCU208, ADC channel‑to‑channel isolation was measured by driving one channel with an external tone frequency and measuring the leakage into adjacent channels. The results show isolation better than 80\thinspace dB, as illustrated in Figure~\ref{fig:crosstalk}, comparable to performance in conventional systems~\cite{Zest}.

\begin{figure}[!htb]
	 	\includegraphics*[width=1\columnwidth]{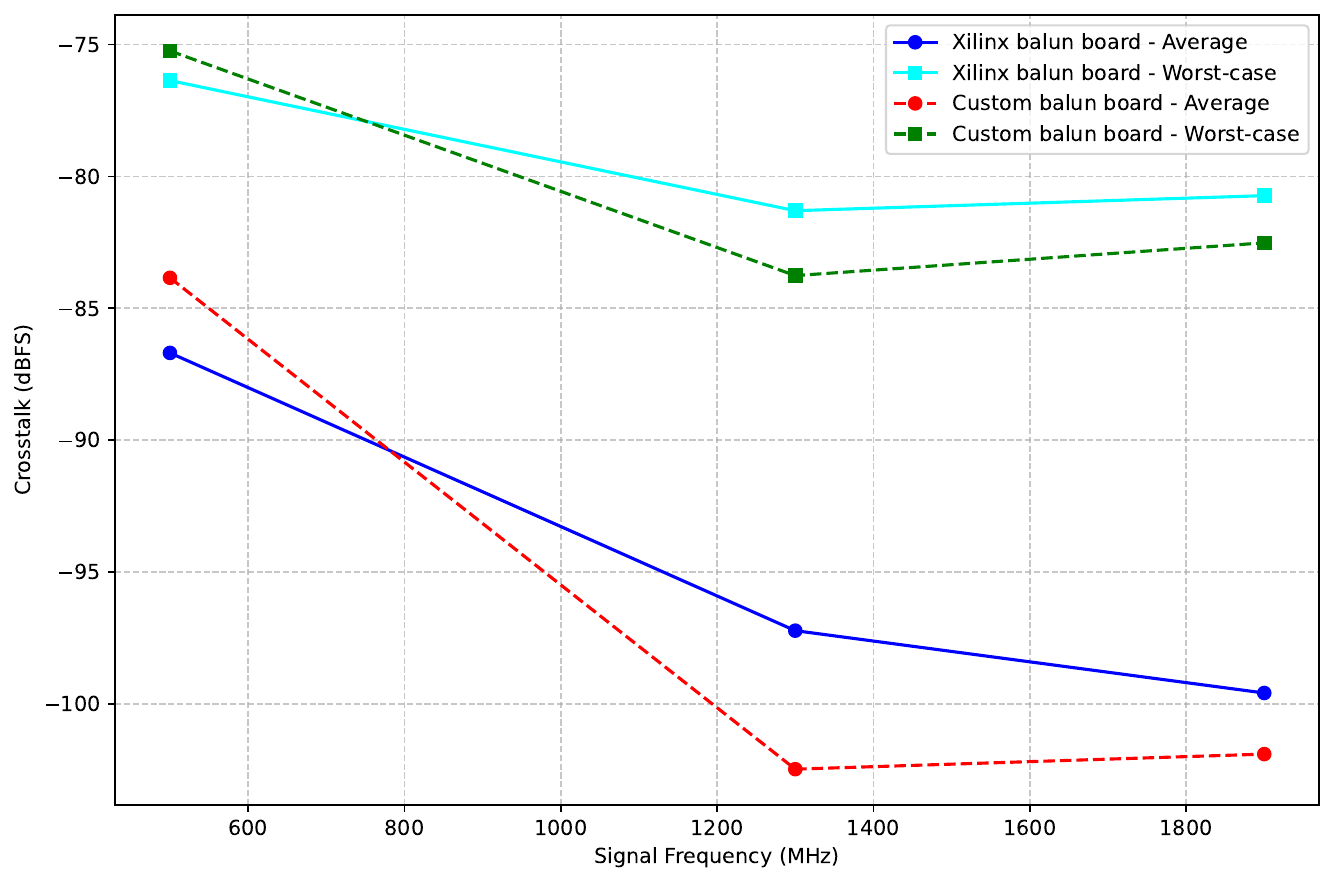}
	 	\caption{Receiver crosstalk comparison between different front-end.}
	 	\label{fig:crosstalk_comp}
	 \end{figure}

	 Crosstalk measurements were also taken at different ADC tone frequencies using Xilinx's stock XM655 board (included with the evaluation kit) and a custom‑built balun board. Figure~\ref{fig:crosstalk_comp} compares the average and worst‑case isolation between the two front‑end boards, with the overall trend showing improved isolation at higher tone frequencies for both configurations.
	 
	 \subsection{Latency}
	 End‑to‑end delay from ADC to DAC using a single channel loopback configuration. The RFSoC direct‑sampling path exhibited a latency of 300\thinspace ns, significantly lower than that of the conventional system~\cite{J-PARC}, due to the elimination of IF stages and associated analog processing delays. Deterministic latency, repeatable delay across resets and power cycles was also verified using the target‑latency configuration, confirming stable and predictable timing behavior.

	 \section{CONCLUSION}
	 This paper has evaluated the performance of the ZCU208 RFSoC‑based system across key LLRF metrics, including phase noise, DAC and ADC characterization, crosstalk, and latency. The results suggest that RFSoC architectures, with their integrated design and advanced clocking capabilities, can provide stability and signal fidelity comparable to conventional designs, while reducing the need for multiple discrete components and associated cabling.

	 When choosing between RFSoC and traditional LLRF architectures, several key factors must be considered. While RFSoC platforms may reduce the number of analog components, the initial investment in RFSoC hardware and the specialized expertise required for development and maintenance can be substantial. Traditional LLRF systems, built from lower‑cost and well‑understood individual components, often involve more complex assembly and calibration processes. Ultimately, the choice depends on application requirements, performance targets, budget, and deployment scale, with ongoing RFSoC advancements expected to further improve their suitability for modern accelerator systems.
	
	 \section{ACKNOWLEDGMENTS}
	 This work was supported by the ALS Project and the Office of Science, Office of Basic Energy Sciences, of the U.S. Department of Energy under Contract No. DE-AC02-05CH11231.

	\ifboolexpr{bool{jacowbiblatex}}%
	{\printbibliography}%
	{%
		
	} 
	%
	%
	
	
\end{document}